\documentclass[prl,twocolumn,preprintnumbers,amsmath,amssymb]{revtex4}

\usepackage{graphicx}% Include figure files
\usepackage{dcolumn}% Align table columns on decimal point
\usepackage{bm}% bold math
\usepackage{color}
\usepackage{epstopdf}

\usepackage{ulem}
\def\eps{\varepsilon}
\def\bk{{\bm \kappa}}

\newcommand{\bX}{ {\bm X}}
\newcommand{\bK}{ {\bm K}}

\def\eps{\varepsilon}
\def\bk{{\bm k}}
\def\bx{{\bm x}}
\def\by{{\bm y}}
\def\bK{{\bm K}}
\def\bX{{\bm X}}
\def\bY{{\bm Y}}

\begin{document}

%\widetext

%\preprint{APS/123-QED}

\title{Coherent soliton states hidden in phase-space \\
and stabilized by gravitational incoherent structures}% Force line breaks with \\

%\author{authors}
\author{Josselin Garnier$^{1}$, Kilian Baudin$^{2}$, Adrien Fusaro$^{2,3}$, 
Antonio Picozzi$^{2}$}
\affiliation{$^{1}$ CMAP, CNRS, Ecole Polytechnique, Institut Polytechnique de Paris, 91128 Palaiseau Cedex, France}
\affiliation{$^{2}$ Laboratoire Interdisciplinaire Carnot de Bourgogne, CNRS, Universit\'e Bourgogne Franche-Comt\'e, Dijon, France}
\affiliation{$^{3}$ CEA, DAM, DIF, F-91297 Arpajon Cedex, France}


\begin{abstract}
We consider the problem of the formation of soliton states from a modulationally unstable initial condition in the framework of the Schr\"odinger-Poisson (or Newton-Schr\"odinger) equation accounting for gravitational interactions. 
We unveil a previously unrecognized regime: 
By increasing the nonlinearity, the system self-organizes into an incoherent localized structure that contains ``hidden" coherent soliton states. 
The solitons are ``hidden" in the sense that they are fully immersed in random wave fluctuations: 
The radius of the soliton is much larger than the correlation radius of the incoherent fluctuations while its peak amplitude is of the same order of such fluctuations. 
Accordingly, the solitons can hardly be identified  in the usual spatial or spectral domains, while their existence is clearly unveiled in the phase-space representation.
Our multi-scale theory based on coupled coherent-incoherent wave turbulence formalisms reveals that the hidden solitons are stabilized and trapped by the incoherent localized structure.
Furthermore, hidden binary soliton systems are identified numerically and described theoretically.
The regime of hidden solitons is of potential interest for self-gravitating Boson models of ``fuzzy" dark matter. It also sheds new light on the quantum-to-classical correspondence with gravitational interactions. 
The hidden solitons can be observed in nonlocal nonlinear optics experiments through the measurement of the spatial spectrogram.
\end{abstract}


\maketitle

Understanding the processes of self-organization in conservative Hamiltonian systems is a difficult problem that has generated significant interest.
For nonintegrable wave systems, the formation of a coherent soliton state plays the role of a ``statistical attractor" for the Hamiltonian system \cite{zakharov88,rumpf_newell_01,jordan_josserand00,ZDP04}.
It is thermodynamically advantageous for the system to generate a large scale soliton, because this allows to increase the amount of disorder (`entropy') in the form of thermalized small scale  fluctuations \cite{zakharov88,rumpf_newell_01,jordan_josserand00,ZDP04,laurie12,nazarenko11,newell_rumpf,Newell01}.

The physical picture becomes more complex when the system exhibits long-range interactions, which dramatically slow down the thermalization process.
A detailed understanding of this process is a subject of growing interest, in relation with peculiar features such as violent relaxation, ergodicity breaking, or inequivalence of thermodynamic ensembles \cite{ruffo_book}.
In this respect, the Schr\"odinger-Poisson equation (SPE) (or Newton-Schr\"odinger equation) appears as a natural theoretical framework to study a wave system with long-range interactions.
The SPE was proposed with the aim of investigating quantum wave function collapse in the presence of a Newtonian gravitational potential \cite{diosi84,penrose96}. 
Actually, the SPE may be obtained as the non-relativistic limit of the self-gravitating Klein-Gordon equation \cite{ruffini69,giulini12}, and thus describes the coupling of classical gravitational fields to quantum matter states.
Soliton solutions of the SPE \cite{chavanis_calmet} have been used to introduce the concept of Bose stars \cite{ruffini69,jetzer92}.
More recently, the SPE has been proposed for a quantum mechanical formulation of dark matter that would solve the `cold dark matter crisis', e.g. the formation of a cusp in the classical description of cold dark matter \cite{chavanis11,suarez14,weinberg15,
witten17,marsh17,braaten19,niemeyer20}.
Indeed, recent 3D numerical simulations of the SPE realized in the cosmological setting remarkably reveal that, as a rule, the system self-organizes into a large scale soliton core, which is surrounded by an incoherent  structure that appears consistent with the classical description 
\cite{schive14a,schive14b,niemeyer16,mocz17,mocz18,bar18,
mocz19,schive20}.
In other words, the repulsive quantum potential (arising from the uncertainty principle) that is inherent to the SPE leads to the formation of a solitonic core that solves the cusp problem of classical cold dark matter.
This Bosonic model for dark matter is known in the literature as fuzzy-dark matter, ultralight axion dark matter, or Bose-Einstein condensate dark matter.

Our aim in this Letter is to unveil a previously unrecognized regime of the SPE. 
Considering a homogeneous initial condition, we show that, by increasing the amount of nonlinearity, the field evolves toward an incoherent localized state that contains `hidden' coherent soliton structures.
The incoherent structure (IS) `hides' coherent soliton states in the following sense: 
(i) The soliton amplitude is of the same order as the fluctuations of the surrounding IS; 
(ii) The radius of the coherent soliton is {\it larger} than the correlation radius of the fluctuations of the IS, but {\it smaller} than the radius of the IS, see Eq.(\ref{eq:sep_scales}).
Then the coherent soliton state can hardly be identified in the usual spatial or spectral domains, while its existence is clearly unveiled in the phase-space representation.
Our theory provides a detailed description of the hidden coherent soliton states, which remarkably reveals that they are trapped and stabilized by the surrounding IS.
Aside from the SPE context, the hidden character of the solitons predicted here has not been discussed before in the soliton literature.

There is a surge of interest in studying analogue gravity phenomena in {\it optical laboratory-based experiments} that recreate some aspects of the full gravitational system \cite{segev15,faccio_bose_star,faccio_lnp,marino19,nazarenko_nse,paredes20}.
% i.e., the study of gravitational effects using artificial systems that recreate some specific aspects of the full gravitational system.
Gravity being inherently nonlinear and nonlocal, the hidden coherent soliton states predicted here could be observed in highly nonlocal nonlinear optics experiments \cite{Segev_rev,kivshar_agrawal03,PR14,rotschild06,cohen06,rotschild08,
vapor,peccianti04,rotschild05,marcucci19}, or alternatively in dipolar Bose-Einstein condensates \cite{baranov08}.

{\it Schr\"odinger-Poisson equation.-} 
We consider a general form of the SPE in spatial dimension $D$:
\begin{eqnarray}
\label{eq:nse_0}
i \partial_{{t}} {\psi} &=& - \frac{\alpha}{2} \nabla^2 {\psi} + {V} {\psi},\\
\label{eq:nse_1}
\nabla^2 {V} &=&  \gamma  \eta_D |\psi|^2,
\end{eqnarray}
where $\alpha > 0$ and $\gamma >0$ are the dispersion and nonlinear coefficients, with $\eta_1= 2$, $\eta_2=2\pi$, $\eta_3=4\pi$.
Accordingly, 
$V =  - \gamma \int  U_D({\bx}-{\by}) |{\psi}(\by)|^2 d\by \equiv -\gamma U_D * |{\psi}|^2$,
with 
$U_1(x) = -|x|$,   $U_2(\bx) = -\log(|\bx|)$,  $U_3(\bx)= 1/|\bx|$.
The SPE describes a Bose gas under 
%the influence of 
its self-induced gravitational potential $V(\bx,t)$ satisfying the Poisson Eq.(\ref{eq:nse_1}) with $\alpha=\hbar/m$ and $\gamma=Gm/\hbar$, where $m$ is the mass of the Bosons and $G$ the Newton gravitational constant.

{\it Hidden soliton regime.-} 
If we denote by $\bar{\rho}$  the typical amplitude of $|\psi|^2$ and by $\ell$ its typical radius, 
%%, i.e. its typical radius of variation.
then $V \sim \gamma \bar{\rho} \ell^2$ and the characteristic nonlinear time scale is $\tau_{nl}=1/(\gamma \bar{\rho}\ell^2)$.
On the other hand, the time scale due to linear dispersion effects is 
$\tau_l = \lambda_c^2/(\alpha/2)$, where $\lambda_c$ is the correlation radius of the field $\psi$.
The healing length 
\begin{eqnarray*}
\xi =  \ell^{-1} \big( \alpha/(2\gamma \bar{\rho})\big)^{1/2}
\end{eqnarray*}
then denotes the spatial scale such that linear and nonlinear effects are of the same order.
The weakly nonlinear (kinetic) regime $\lambda_c \ll \xi$  (or $\tau_l/\tau_{nl} = \lambda_c^2/\xi^2 \ll 1$) is described by the recently developed wave turbulence (WT) kinetic theory \cite{nazarenko_nse,levkov18}. 
This is {\it not} the regime addressed in this work.

It proves convenient to normalize the healing length $\tilde{\xi} = \xi / \Lambda$ with respect to the Jeans length $\Lambda = \big( \alpha/(2\gamma {\bar \rho}) \big)^{1/4}$, which denotes the cut-off spatial length below which a homogeneous wave is modulationally stable.
The dimensionless parameter ${\tilde \xi}=\Lambda/\ell=(\xi/\ell)^{1/2}$
is directly related to a parameter $\Xi=(\hbar/m)^2/(2 \ell^4 {\bar \rho} G) =
\tilde{\xi}^4$ that has been  shown to control the quantum to classical limit, i.e. the 
 Schr\"odinger-Poisson to Vlasov-Poisson correspondence in the limit $\hbar/m \to 0$ \cite{mocz18}. 
Indeed, for ${\tilde \xi} \lesssim 1$, the radius $\ell$ of a gravitational structure is of the same order as the healing length $\xi \sim \ell$, so that linear `quantum effects' play a fundamental role and the system exhibits a {\it coherent} dynamics that is essentially dominated by soliton structures. 
Massive numerical simulations in the cosmological setting have widely explored this regime \cite{niemeyer20,schive14a,schive14b,niemeyer16,mocz17,bar18,mocz18,mocz19,schive20}: 
They show the formation of an IS that is dominated in its center by a large amplitude coherent soliton peak $\rho_S$, typically much larger than the average density of the surrounding IS, $\rho_S \gg {\bar \rho}_{IS}$, see \cite{bar18}.
Furthermore, the soliton radius $R_S$ is typically of the order of the correlation radius of the fluctuations of the IS, $R_S \sim \lambda_c$ \cite{schive14a,schive14b,niemeyer16,mocz17,mocz18,mocz19,niemeyer20b}.

On the other hand, in the strongly nonlinear regime ${\tilde \xi} \ll 1$, the dynamics is dominated by the gravitational interaction described by the Vlasov-Poisson equation (VPE), which is a  kinetic equation inherently unable to describe coherent soliton structures.
In other words, in the 
%strongly nonlinear 
regime ${\tilde \xi} \ll 1$ where $\alpha/\gamma \propto (\hbar/m)^2 \to 0$, coherent solitons should gradually disappear and the field $\psi(\bx,t)$ should exhibit a purely incoherent dynamics featured by the generation of a large scale IS \cite{mocz18}.
The main result of our work is to show that such an IS is not purely incoherent, but still contains `hidden' soliton states: The IS with typical average density ${\bar \rho}_{IS}$, radius $\ell$ and correlation radius $\lambda_c$, helps stabilizing a soliton of amplitude $\rho_S$ and typical radius $R_S$ verifying \cite{supplement}: 
\begin{eqnarray}
 \lambda_c \sim \xi \ll R_S \sim \Lambda \ll \ell, \quad  \rho_S \sim {\bar \rho}_{IS} .
\label{eq:sep_scales}
\end{eqnarray}
More precisely, $R_S  \sim \Lambda=\sqrt{\xi \ell}$ is the geometric average of $\ell$ and $\xi$. 
Note that the correlation radius is of the order of the de Broglie wavelength, $\lambda_c \sim \lambda_{\rm dB}$, where $\lambda_{\rm dB} \to 0$ in  the quantum-to-classical (SPE to VPE) limit  \cite{mocz18}.

\medskip

{\it Simulations.-} 
An example of the regime (\ref{eq:sep_scales}) is illustrated in Fig.~1.
We consider SPE simulations in 1D because the parameter ${\tilde \xi}$ (or $\Xi$) does not depend on the spatial dimension $D$.
The advantage with respect to 3D simulations is that much smaller values of the parameter $\Xi$ can be reached in 1D.
In Fig.~1 we consider $\Xi \simeq 5\times 10^{-8}$, a  value that appears inaccessible in 3D, where %typically 
$\Xi > 10^{-4}$ \cite{mocz17,mocz18,schive20}.
In other words, the novel regime (\ref{eq:sep_scales}) seems out of reach of current 3D simulations \cite{mocz18}.

The initial condition in Fig.~1 is a homogeneous wave $\psi(x,t=0)=\sqrt{{\bar \rho}}$ with a superimposed small noise to initiate the modulational (gravitational) instability.
The instability is followed by a gravitational collapse, which is regularized by the formation of a virialized IS 
\cite{schive14a,schive14b,niemeyer16,mocz17,mocz18,bar18,mocz19}.
Note that the localized IS exhibits properties similar to those of incoherent optical solitons in nonlocal nonlinear media \cite{cohen06,rotschild08,PR14}.
The IS in Fig.~1(a) does not exhibit apparent coherent soliton structures (also see Movie~1 in \cite{supplement}).
This appears consistent with the SPE to classical VPE correspondence.
Unexpectedly, however, the IS is not purely incoherent, but contains hidden coherent soliton structures.
Such soliton entities are unveiled by a phase-space analysis of the field $\psi(x)$ provided by the Husimi representation (smoothed  Wigner transform) \cite{mocz18,supplement},  
which denotes the field spectrum at different spatial positions.
In optics, the Husimi transform is provided by the measurement of the spectrogram \cite{waller12}.
In phase-space, the solitons are characterized by high intensity spots, while the surrounding small amplitude fluctuations denote the IS, see Fig.~1. 
The coherent soliton has a spectral width $\Delta k_{S}$ much smaller than the spectral width of the IS, $\Delta k_{IS} \gg \Delta k_{S}$, which means that the radius of the soliton, $R_S \sim 2\pi/\Delta k_{S}$, is much larger than the correlation radius $\lambda_{c} \sim 2\pi/\Delta k_{IS}$ of the IS.
Remarking furthermore that $R_S \ll \ell$, 
we can clearly observe the separation of spatial scales (\ref{eq:sep_scales}) in Fig.~1(e).

The SPE simulations remarkably show that the hidden solitons get {\it trapped} by the IS, as revealed by the phase-space dynamics in Fig.~1.
In particular, the untrapped soliton labelled ``5" in Fig.~1(b) is not robust and disappears at $t \simeq 27 \tau$, see Movie~1 in \cite{supplement}.
Then the IS plays the role of an effective trapping potential for a soliton, as will be confirmed by the theory, see Eq.(\ref{eq:nls_A0}).
The solitons hidden within the IS exhibit  complex dynamics.
Two solitons can spin around each other in phase-space thus forming a {\it binary system},  see Fig.~2 and Movie~2 \cite{supplement}.
The number of solitons decreases with time, eventually leading to a single soliton that exhibits an ellipsoidal periodic motion in phase-space, see Fig.~3 and Movie~3 \cite{supplement}.

\begin{center}
\begin{figure}
\includegraphics[width=1\columnwidth]{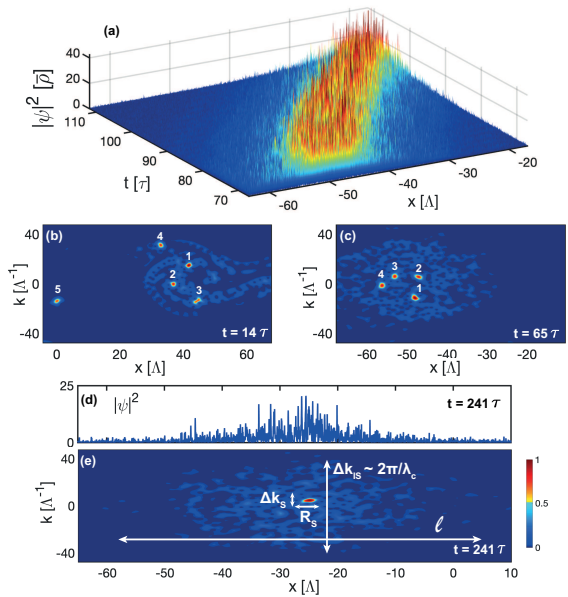}
\caption{
\baselineskip 10pt
{\bf Unveiling coherent solitons in phase-space:}
SPE simulation for ${\tilde \xi} \simeq 1.5 \times 10^{-2}$:
(a) Spatio-temporal evolution of the density $|\psi|^2(x,t)$. 
(b)-(c)-(e) The hidden solitons are unveiled in phase-space 
by high intensity spots (labels (1)-(5)).
The number of solitons decreases with time, eventually leading to a single soliton (e).
Density $|\psi|^2(x)$ (d) and corresponding phase-space portrait (e) at $t=241\tau$,
showing the separation of the three spatial scales $\lambda_c \sim 2\pi/\Delta k_{IS} \ll R_S \sim 2\pi/\Delta k_{S} \ll \ell$, see Eq.(\ref{eq:sep_scales}).
Parameters: $D=1$ for $x \in [-L/2,L/2]$ with periodic boundary conditions ($L=135 \Lambda$, $\tau=2 \Lambda^2/\alpha$), see \cite{supplement} and Movie~1.
}
\end{figure}
\end{center}

{\it Effective Schr\"odinger-Poisson equation.-}
We develop the theory in the general WT framework \cite{zakharov92,laurie12,Newell01,ZDP04,nazarenko11,
newell_rumpf,PR14,PRL11,Newell13,NC15,PD16,Turitsyn_rev}. 
%The WT theory has been shown to provide a natural asymptotic closure of the hierarchy of moment equations for a system of weakly nonlinear dispersive waves \cite{}.   
Because $\lambda_c \sim \xi$ (see Eq.(\ref{eq:sep_scales})), the IS does {\it not} evolve in the weakly nonlinear regime \cite{supplement}.
It will be described by a WT-VPE that generalizes to long-range interactions \cite{PR14} the WT Vlasov equation describing random waves in optics \cite{PR14,Segev_rev}, hydrodynamics \cite{Newell13,Onorato13} or plasmas \cite{lvov77,ZakharovPR85,nazarenko92}.

We describe the coupled coherent-incoherent dynamics of the soliton immersed in the IS by deriving a coupled system of SPE and WT-VPE.
The soliton is characterized by a non-vanishing average $\left< \psi \right> \neq 0$, so that the field can be decomposed into a coherent component $A(\bx,t)$ and an incoherent component 
$\phi(\bx,t)$ of zero mean ($A=\left<\psi\right>, \phi =\psi - \left< \psi \right>$):
\begin{eqnarray*}
\psi(\bx,t)=A(\bx,t)+\phi(\bx,t).
%\label{eq:decomposepsi}
\end{eqnarray*}
The local spectrum of the IS is the {\it average Wigner transform}
$n({\bm k},{\bm x},t)= \int  \left< \phi(\bx+{\bm y}/2,t)  \phi^*(\bx-{\bm y}/2,t)\right>  \exp(-i {\bm k} \cdot {\bm y}) \, d{\bm y}$.
Starting from the SPE (\ref{eq:nse_0}-\ref{eq:nse_1}), we obtain the result that the soliton and IS components are governed by the coupled SPE and WT-VPE \cite{supplement}:
\begin{eqnarray}
&& i\partial_t A  = - \frac{\alpha}{2} \nabla^2 A + A V , 
\label{eq:nls_0}
\\
&& \partial_t n({\bm k},{\bm x}) +\alpha {\bm k}\cdot \partial_{\bx} n({\bm k},{\bm x}) - \partial_{\bm x} V \cdot \partial_{\bm k} n({\bm k},{\bm x}) = 0 .
\quad \quad 
\label{eq:vlasov_0}
\end{eqnarray}
Eqs.(\ref{eq:nls_0}-\ref{eq:vlasov_0}) are coupled by the potential $V({\bm x},t) = - \gamma U_D * ( |A|^2+ { \rho}_{IS})$, which is the sum of the coherent and incoherent contributions with 
$\rho_{IS}({\bm x},t)=\left< |\phi(\bx,t)|^2 \right>=(2 \pi)^{-D}  \int n({\bm k},{\bm x},t) d \bk$ the average density of the IS.

Further insight into the coupled SPE and WT-VPE (\ref{eq:nls_0}-\ref{eq:vlasov_0}) is obtained through a multi-scale expansion in the small parameter $\eps \equiv {\tilde \xi}\ll 1$:
$A(\bx,t) = A^{(0)}(\bx,t)$, $n(\bk,\bx,t) = \eps^D n^{(0)}(\eps \bk, \eps \bx, t)$.
This scaling gives Eq.(\ref{eq:sep_scales}): $\lambda_c/\Lambda =O(\varepsilon)$, $\ell / \Lambda = O(\varepsilon^{-1})$, $R_S/\Lambda =O(1)$, $\rho_S \sim |A|^2 \sim O(1)$, $n \sim O( \varepsilon^D)$ and $\rho_{IS}\sim O(1)$.
Accordingly, we derive an effective SPE (ESPE) for the coherent component \cite{supplement}: 
\begin{eqnarray}
 i\partial_t A  = - \frac{\alpha}{2} \nabla^2  A + V_{S} A + \gamma  q_D   \rho_{0} |\bx|^2 A,
\label{eq:nls_A0} 
\end{eqnarray}
where $V_{S}(\bx,t)=-\gamma U_D * |A|^2$, 
${\rho}_0(t) = {\rho}_{IS}(\bx={\bf 0},t)$ is the central average density of the IS, 
and $q_D$ depends on the dimension, $q_1=1,  q_2=\pi/2, q_3=2\pi/3$.
The ESPE (\ref{eq:nls_A0}) reveals that the coherent (soliton) component experiences its self-gravitational potential $V_{S}$ and an unexpected parabolic trapping potential due to the IS.

{\it Dynamics of hidden solitons in $D$ dimension.-}
We describe the general form of the spinning binary soliton by using the variational approach (for $D=1,3$).
We consider the Lagrangian of the ESPE (\ref{eq:nls_A0}) with the Gaussian ansatz:
\begin{eqnarray*}
\nonumber
A(\bx,t) = \sum_{j=1}^2 a_j(t) \exp \Big( -\frac{|\bx-\bx_{o,j}(t)|^2}{2R_{S,j}^2(t)} +i \Phi_j(\bx,t) \Big)  ,
\end{eqnarray*}
where $\Phi_j(\bx,t) =  \bk_{o,j}(t) \cdot (\bx-\bx_{o,j}(t))+b_j(t)|\bx-\bx_{o,j}(t)|^2 +\nu_j(t)$. 
The evolution of the phase-space coordinates of the $j-$th soliton $(\bx_{o,j}(t),\bk_{o,j}(t))$ are obtained from the principle of least action through the Euler-Lagrange equations \cite{supplement}.

\begin{center}
\begin{figure}
\includegraphics[width=1\columnwidth]{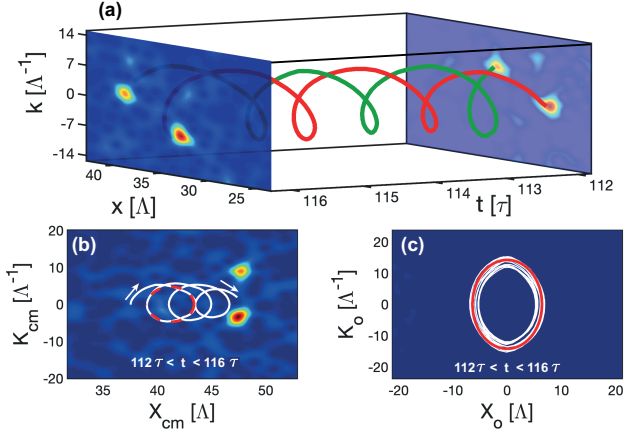}
\caption{
\baselineskip 10pt
{\bf Hidden binary soliton:}
(a) SPE simulation reported in Fig.~1 (at longer time)
%SPE simulation for the same parameters as in Fig.~1 
showing two solitons that orbit around each other in phase-space.
(b) The center of mass exhibits an ellipsoidal motion with period $\tau_{\rm cm}^{\rm num} \simeq 1.56 \tau$ in agreement with the theory, see Eq.(\ref{eq:tau_cm}) [the horizontal shift is due to the motion of the IS, see Fig.~1(a)].
The dashed red line reports the theoretical ellipse $H_{\rm cm}$ from Eq.(\ref{eq:E_cm}).
(c) The spinning period of the solitons around each other $\tau_{\rm bin}^{\rm num} \simeq 1.43 \tau$ is in agreement with the theory, see Eq.(\ref{eq:tau_bin}).
The red line reports the theoretical prediction $H_{o}$ from Eq.(\ref{eq:E_bin}).
See Movie~2 in \cite{supplement}.
}
\end{figure}
\end{center}

{\it Ellipsoidal motion of the center of mass.-}
The dynamics of the binary soliton can be decomposed into the motion of the center of mass (CM) and the mutual relative displacement of the two solitons in the CM reference frame. 
The equations for the CM,  $\bX_{\rm cm} = (M_{S,1} \bx_{o,1}+M_{S,2} \bx_{o,2})/(M_{S,1}+M_{S,2})$, and $\bK_{\rm cm} = (M_{S,1} \bk_{o,1}+M_{S,2} \bk_{o,2})/(M_{S,1}+M_{S,2})$, can be recast in Hamiltonian form $\partial_t \bX_{\rm cm} =  \partial_{\bK_{\rm cm}} H_{\rm cm}$, $\partial_t \bK_{\rm cm} =  - \partial_{\bX_{\rm cm}} H_{\rm cm}$ with 
\begin{eqnarray}
H_{\rm cm}=q_D\gamma \rho_0 |\bX_{\rm cm}|^2 + \frac{\alpha}{2} |\bK_{\rm cm}|^2.
\label{eq:E_cm}
\end{eqnarray}
The barycenter of the binary soliton then exhibits a periodic ellipsoidal motion in phase-space with a revolution period 
\begin{eqnarray}
\tau_{\rm cm}= \sqrt{2} \pi/\sqrt{\alpha \gamma q_D \rho_0}.
\label{eq:tau_cm}
\end{eqnarray}

We first comment the case $D=1$ through the SPE simulation reported in Fig.~2: The average density $\rho_0 = (3.9 \pm 0.2) {\bar \rho}$ 
gives $\tau_{\rm cm}=(1.56 \pm 0.04) \tau$ from (\ref{eq:tau_cm}), which is in agreement with the simulation ($\tau_{\rm cm}^{\rm num} \simeq 1.56 \tau$).

The revolution period (\ref{eq:tau_cm}) also applies to the ellipsoidal motion of a {\it single soliton} where the CM coincides with the soliton position.
In  Fig.~3, $\rho_0 = (4.2\pm 0.2) {\bar \rho}$ gives $\tau_{\rm cm}=(1.52 \pm 0.04) \tau$, which is in agreement with the SPE simulation ($\tau_{\rm cm}^{\rm num} \simeq 1.52 \tau$) \cite{supplement}.

\smallskip
For $D=3$, the dynamics $\bX_{\rm cm}(t)$ lies in a plane and exhibits an {\it ellipsoidal} motion:
%In polar coordinates, it has the form 
$\bX_{\rm cm}= ({\cal R}(\theta) \cos \theta, {\cal R}(\theta) \sin \theta, 0)$
where ${\cal R}(\theta) = (w_- \cos^2\theta+w_+\sin^2 \theta)^{-1/2} c_1^{-1/4}$
with $\theta(t) = {\rm arctan}\big( w_- \tan( c_o \sqrt{c_1} t)\big)$, 
$c_1= 2 q_3 \alpha\gamma \rho_0/c_o^2$,  $w_{\pm}$ and $c_{o,1}$ being constants of the motion \cite{supplement}.

%\begin{widetext}
\begin{center}
\begin{figure}
\includegraphics[width=1\columnwidth]{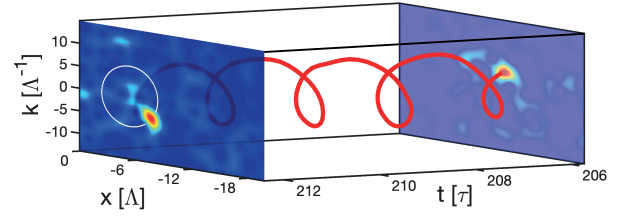}
\caption{
\baselineskip 10pt
{\bf Hidden single soliton:}
SPE simulation reported in Fig.~1 (at longer time)
% for the same parameters as in Fig.~1 
showing the phase-space evolution of a single soliton.
The soliton exhibits an ellipsoidal phase-space motion with period $\tau_{\rm cm}^{\rm num} \simeq 1.52 \tau$ in agreement with the theory, see Eq.(\ref{eq:tau_cm}).
The white line reports the theoretical ellipse $H_{\rm cm}$ predicted in Eq.(\ref{eq:E_cm}).
See Movie~3 in \cite{supplement}.
}
\end{figure}
\end{center}
%\end{widetext}

{\it Revolution period for the binary soliton.-}
The Hamiltonian equations governing the relative position of the binary soliton, namely $\bX_o = \bx_{o,1}-\bx_{o,2}$, $\bK_o = \bk_{o,1}-\bk_{o,2}$, read 
$\partial_t \bX_o = \partial_{\bK_o} H_o$, $\partial_t \bK_o = -\partial_{\bX_o} H_o$, with 
\begin{eqnarray}
 H_{o} \! = \! q_D\gamma \rho_0 |\bX_o|^2 + 
 \! \gamma (M_{S,1} \! + \! M_{S,2} ) 
|\bX_o|^{2-D} \!
+ \! \frac{\alpha}{2}|\bK_o|^2 .
\label{eq:E_bin}
\end{eqnarray}
For $D=1$, the phase-space trajectory is reported in Fig.~2.
The spinning binary soliton exhibits a revolution period \cite{supplement}:
\begin{equation}
{\tau}_{\rm bin} =
4\sqrt{2} \ {\rm arcsin}
( \sqrt{ \beta/2}) / \sqrt{\alpha \gamma \rho_0},
\label{eq:tau_bin}
\end{equation}
where $\beta = 1 -(M_{S,1} +M_{S,2} )/\sqrt{\chi}$, and $\chi=(M_{S,1} +M_{S,2} ) ^2 +4 \rho_0 d(\rho_0 d + M_{S,1} +M_{S,2})$, with $d$ the maximal soliton distance.
The spinning for the binary soliton is always faster than for a single soliton ($\tau_{\rm bin} < \tau_{\rm cm}$), as confirmed by the SPE simulation in Fig.~2 where $\tau_{\rm bin} = (1.43 \pm 0.04) \tau$ is in agreement with the simulation ($\tau_{\rm bin}^{\rm num} \simeq 1.43 \tau$).

For $D=3$, the motion of the binary soliton lies in a plane: $\bX_o = ({\cal R}(\theta) \cos \theta, {\cal R}(\theta) \sin \theta, 0)$,
where $u(\theta) = 1/{\cal R}(\theta)$ is solution of $\partial_\theta^2 u +u = c_1/u^3+c_2$, 
where 
$\partial_t \theta = c_o {\cal R}(\theta)^{-2}$, $c_{o,1,2}$ being constants of the motion.
The orbit $\bX_o$ is not closed in general and the motion in the plane exhibits a perihelion precession \cite{supplement}.

{\it Discussion and perspectives.-}
We have reported a novel regime of the SPE characterized by hidden soliton states that are trapped and stabilized by the IS.
The  regime of hidden solitons can be observed in highly nonlocal nonlinear optics experiments with long-range thermal nonlinearities, in line with the recent emulations of rotating Bose stars, or gravitational lensing and red-shifts \cite{segev15,faccio_bose_star}.
%The phase-space distribution can be retrieved experimentally from a windowed 
The hidden solitons can be unveiled experimentally through the measurement of the optical spectrogram \cite{waller12,supplement}.

Aside from its relevance to Bosonic models of fuzzy dark matter, 
our work sheds new light 
on the quantum-to-classical (or SPE to VPE) correspondence in the limit $\hbar/m \to 0$: 
The hidden solitons revealed here 
refer to the {\it latest residual quantum correction} preceding the purely classical limit provided by the VPE.

{\it Acknowledgements.-}   
The authors are grateful to J. Niemeyer for drawing our attention to this problem and for the fruitful discussions and suggestions during the early stage of this work.
The authors also thank S. Rica and S. Nazarenko for valuable comments.
We acknowledge financial support from the French ANR under Grant No. ANR-19-CE46-0007 (project ICCI), French program ``Investissement d'Avenir," Project No. ISITE-BFC-299 (ANR-15 IDEX-0003); H2020 Marie Sklodowska-Curie Actions (MSCA-COFUND) (MULTIPLY Project No. 713694).

%%%%%%%%%%%%%%%%%%%%%%%%%%%%%%%%%%%%%%%%%%%%%%%%%%%%%
%%%%%%%%%%%%%%%%%%%%%%%%%%%%%%%%%%%%%%%%%  Appendix
%%%%%%%%%%%%%%%%%%%%%%%%%%%%%%%%%%%%%%%%%%%%%%%%%%%%%
%{\bf Supplementary Material}

%\appendix 

\begin{widetext}

\section{Supplementary Material}

\section{I. Numerical techniques}

\subsection{SPE in a finite box for simulations}
%\label{app:A}%
The numerical simulations are carried out on a system set in a bounded domain with prescribed boundary conditions:
\begin{align}
\label{eq3}
i \partial_{{t}} {\psi} + \frac{\alpha}{2} \Delta_{{\bx}} {\psi} - {V}_L {\psi}  =0 , \\
\label{eq4}
 {V}_L = - \gamma \eta_D U_{D,L} * |\psi|^2 ,
\end{align}
where $\bx \in [-L/2,L/2]^D$ with periodic boundary conditions.
Here $U_{D,L}(\bx) $ is the periodic function equal to $U_D(\bx) -c_{D,L}$ in $[-L/2,L/2]^D$ and 
$c_{D,L} =  L^{-D} \int_{[-L/2,L/2]^D} U_D(\bx) d\bx$ so that $U_{D,L}$ has mean zero. 
Eq.~(\ref{eq4}) is equivalent to 
%\begin{equation}
${V}_L = - \gamma \eta_D U_{D} * ( |\psi|^2 - \bar{\rho}_L)$,
% \end{equation}
with $\bar{\rho}_L= L^{-D} \int_{[-L/2,L/2]^D} |\psi(\bx)|^2 d\bx$,
and $V_L$ in the simulations has zero mean.
% and the corresponding nonlinear energy 
%${\cal H}_{nl,L}= \frac{1}{2}\int {V}_L(\bx,t) |\psi|^2(\bx,t) d\bx$
%is reported in Figs.~\ref{fig:sol_regim_1d}, \ref{fig:inc_sol_regim_1d}, \ref{fig:IS_CS_2d} with this potential. 
%differs from the nonlinear energy in open medium ${\cal H}_{nl}=\frac{1}{2}\int {V}(\bx,t) |\psi|^2(\bx,t) d\bx$ by a time-independent constant ${\cal H}_{nl,L}= {\cal H}_{nl} +\gamma {\cal M}^2 c_{D,L}/2$, ${\cal M}= \int_{[-L/2,L/2]^D} |\psi(\bx)|^2 d\bx$.
The potential $V_L$ is numerically computed by Fourier transform \cite{mocz17}.
Note that the formulation Eqs.(1-2)  in the open medium (used for the theoretical analysis) and (\ref{eq3}-\ref{eq4}) in the periodic medium (used for the numerical analysis) have an apparent departure in the definitions of the potential $U$ which differ by a constant. 
%However, the addition of a constant to the potential $U$ changes $\psi$ only through a simple phase term that is linear in time, which does not modify the dynamics.
However, the substraction of the constant $c_{D,L}$ to $U_D$ can be removed by multiplying $\psi$ by $\exp( -i  \gamma \eta_D c_{D,L} {M} t)$, with ${M}= \int_{[-L/2,L/2]^D} |\psi(\bx)|^2 d\bx = L^D \bar{\rho}_L$.
% which does not modify the dynamics.

%\smallskip
%\noindent
%\blue{\sout{{\bf Normalized healing length:} The parameter $\tilde{\xi}  \simeq 7.4\times 10^{-3}$ is computed with $\ell = L = 135 \Lambda$ in the simulations of Figs.~1-3.}} 

\smallskip
\noindent
{\bf Jeans length:} The growth-rate of the gravitational instability in $\sigma_{\rm inst} = \sqrt{\alpha \gamma {\bar \rho}-\alpha^2 k^4/4}$, so that modes with $k \ge k_J=2\pi/\Lambda_J$ with $\Lambda_J=\sqrt{2}\pi (\alpha/(\gamma {\bar \rho}))^{1/4}$ are stable. $\Lambda=(\alpha/(2 \gamma {\bar \rho}))^{1/4} \simeq \Lambda_J$ is the normalizing length scale used in the simulations in Figs.~1-3.
We solved numerically the dimensionless SPE: $i \partial_{\tilde{t}}\tilde{\psi}  =- \nabla_{\tilde \bx}^2 \tilde{\psi} -  \tilde{\psi} {\tilde U}_{D,L} * |\tilde{\psi}|^2$, with $\tilde{\bx}=\bx/\Lambda$, $\tilde{t}= t/\tau $, $\tilde{\psi}=\psi/ \sqrt{{\bar \rho}}$, $\tau= 2\Lambda^2/\alpha$ and ${\tilde L}=L/\Lambda=135$ in Figs.~1-3.

\smallskip
\noindent
{\bf De Broglie wavelength:} It is defined by $\lambda_{\rm dB}=h/(mv)$, where $v$ is the `hydrodynamic' velocity defined from the gradient of the phase $\varphi$ of $\psi=\sqrt{{\rho}}\exp(i \varphi)$. With our notations ${\bm v}=\alpha \nabla \varphi$, so that $\lambda_{\rm dB} = 2 \pi / |\nabla \phi|  \sim  2 \pi \lambda_c$  where $\lambda_c$ is the correlation radius of $\psi$.

\subsection{Wigner and Husimi transforms, and the optical spectrogram}
%\label{app:B}%
The empirical Wigner  transform 
\begin{align}
W(\bk,\bx,t)= \int \psi(\bx+\by/2,t)  \psi^*(\bx-\by/2,t)\exp(-i {\bm k} \cdot\by) \, d\by,
\end{align}
is not statistically stable (its standard deviation is larger than its statistical average).
It is necessary to smooth it with respect to  $\bk$ and $\bx$ to get a statistically stable quantity.
We consider the smoothed Wigner transform or Husimi function
\begin{align}
W_\sigma (\bk,\bx,t)= \frac{1}{\pi^2} \iint W(\bk',\bx',t)
\exp\Big( - \frac{|\bx-\bx'|^2}{\sigma^2} - \sigma^2 |\bk-\bk'|^2\Big) 
d\bk' d\bx'   ,
\label{eq:smoothw1}
\end{align}
which can also be written as
\begin{align}
W_\sigma (\bk,\bx,t)= \frac{1}{\pi \sigma^2}
\Big| \int \psi(\by) \exp\Big(-\frac{|\bx-\by|^2}{2\sigma^2} \Big)
\exp( - i \bk \cdot \by) d\by  \Big|^2   .
\label{eq:smoothw2}
\end{align}
This quantity is statistically stable for all $\sigma>0$. 
When the field is fully incoherent 
the standard deviation of $W_\sigma$ is equal to its average,
more exactly, $W_\sigma$ follows an exponential distribution because it is the square modulus a circular complex Gaussian random variable by (\ref{eq:smoothw2}). 
A small (resp. large) $\sigma$ means that the smoothing is smaller (resp. larger) in $\bx$ than in $\bk$  by (\ref{eq:smoothw1}).
A good trade-off (with equal smoothing in $\bx$ and $\bk$) is achieved by choosing $\sigma \simeq\sqrt{\Delta x/ \Delta k}$ when the radius of $n$ is $\Delta k$ in $\bk$ and $\Delta x$ in $\bx$.
%In Fig.~1 we have $\Delta k \sim 66 \Lambda^{-1}$ and $\Delta x \sim 36 \Lambda$ and we have chosen $\sigma =  0.74 \Lambda$.

\smallskip
\noindent
{\bf Relation with the optical spectrogram:} The measurement of the Husimi transform is known in optics as the `spectrogram'.
It is measured by performing a windowed Fourier transform by passing the beam through a small aperture (Gaussian in (\ref{eq:smoothw2})), whose spatial position is scanned along the optical beam \cite{waller12}.
%The local spectrum in the main text is  $n(\bk,\bx,t)=\left<W(\bk,\bx,t)\right>$.
%It is measured by performing a windowed Fourier transform: The light that passes through a small aperture is optically Fourier-transformed onto a camera. 
%The aperture (Gaussian in (\ref{eq:smoothw2})) is scanned in space while capturing the Fourier domain intensity for each position of the optical beam \cite{waller12}. 
%The local spectrum defined in the main text refers to the averaged Wigner transform $n(\bk,\bx,t)=\left<W(\bk,\bx,t)\right>$.

\section{II. Derivation of the effective SPE, Eq.(6) }

\subsection{System of coupled SPE and WT-VPE, Eqs.(4-5)}

The hidden solitons are characterized by a non-vanishing average $\left< \psi \right> \neq 0$, so that we decompose the field into a coherent component $A(\bx,t)$ and an incoherent component 
$\phi(\bx,t)$ of zero mean ($A=\left<A\right> \neq 0, \left< \phi \right> =0$):
%that exhibits Gaussian statistics (justified by the long-range regime), 
\begin{eqnarray}
\psi(\bx,t)=A(\bx,t)+\phi(\bx,t).
\label{eq:decomposepsi}
\end{eqnarray}
Following the usual procedure \cite{PR14}, we define the spectrum of the IS as 
$n({\bm k},{\bm x},t)= \int  C(\bx,\by,t)  \, \exp(-i {\bm k} \cdot \by) \, d \by$, where the correlation function 
%\begin{eqnarray}
$C(\bx,\by,t)=\left< \phi(\bx+\by/2,t)  \phi^*(\bx-\by/2,t)\right>$,
%\end{eqnarray}
is defined from an average over the realizations $\left< \cdot \right>$.
Starting from the SPE (1-2), we obtain Eqs.(4-5) (main text):
\begin{eqnarray}
&& i\partial_t A  = - \frac{\alpha}{2} \nabla^2 A + A V , 
\label{eq:nls_0}
\\
&& \partial_t n({\bm k},{\bm x}) +\alpha {\bm k}\cdot \partial_{\bx} n({\bm k},{\bm x}) - \partial_{\bm x} V \cdot \partial_{\bm k} n({\bm k},{\bm x}) = 0 ,
\quad \quad 
\label{eq:vlasov_0}
\end{eqnarray}
which are coupled to each other by the {\it averaged} long-range gravitational potential 
\begin{eqnarray}
%&& V({\bm x},t) = - \gamma \int U_d({\bm x}-{\bm y}) \ \big( |A|^2({\bm y},t) + \rho({\bm y},t) \big) \ d{\bm y}, \quad \quad 
&& V({\bm x},t) = - \gamma \int U_D({\bm x}-{\bm y})  \big( |A|^2({\bm y},t) + \rho_{IS}({\bm y},t) \big) d{\bm y}, \quad \quad 
\label{eq:V_0} \\
&& \rho_{IS}({\bm x},t)=\left< |\phi(\bx,t)|^2 \right>=\frac{1}{(2 \pi)^D}  \int n({\bm k},{\bm x},t) d \bk. \quad \quad
\label{eq:N_0} 
\end{eqnarray}
%$\rho({\bm x},t)$ denotes the {\it average density} of the IS, which depends on the spatial variable $\bx$ because the IS exhibits fluctuations that are not homogeneous in space.
% -- it should not be confused with the density $\rho(\bx,t)=|\psi(\bx,t)|^2$ (see e.g. Fig.~\ref{fig:NLSvsVlasov} where $\rho(\bx)$ and $\rho({\bm x})$ are superposed).
%Both the mass of the IS $M_{\rm IS}=\int \rho(\bx,t) d\bx$ and the mass of the CS $M_S=\int |A(\bx,t)|^2 d\bx$ are conserved, with the total mass ${\cal M}=M_S+M_{\rm IS}$.
{\bf Validity of the WT-VPE:} The WT-VPE (\ref{eq:vlasov_0}) is valid beyond the weakly nonlinear regime \cite{PR14}.
Thanks to the long-range nature of the interaction, the system exhibits a self-averaging
property of the nonlinear response, $ \int  U_D({\bx}-{\by}) |{\phi}(\by)|^2 d\by \simeq \int  U_D({\bx}-{\by}) \left<|{\phi}(\by)|^2\right> d\by$. 
Substitution of this property into the SPE leads to an automatic closure of the hierarchy of the moment equations. 
Using statistical arguments similar to those in Ref.\cite{garnier03}, one can show that, owing to a highly nonlocal response, the statistics of the incoherent wave turns out to be Gaussian.

\smallskip
\noindent
{\bf Distinction with the Boltzmann VPE:} It is important to distinguish the WT-VPE (\ref{eq:vlasov_0}) from the collisionless Boltzmann VPE \cite{mocz18,uhlemann14}: At variance with the VPE describing a spiky distribution, the WT-VPE describes the smooth evolution of the second-order moment $n(\bk,\bx)$ defined from the {\it average} over the realizations $\left< \cdot \right>$.

\subsection{Multi-scale expansion theory to derive the effective SPE [Eq.(6)]}

The partially coherent field $\psi$ is of the form (\ref{eq:decomposepsi}). 
It is composed of a coherent soliton $A(\bx,t)$ and an IS characterized by its spectrum $n(\bk,\bx,t)$.
They satisfy the coupled equations (\ref{eq:nls_0}-\ref{eq:vlasov_0}) 
with the long-range potential (\ref{eq:V_0}).
The average density $\rho_{IS}(\bx,t)$ can be expressed in terms of the spectrum $n$ as (\ref{eq:N_0}).
We show that the scaling regime $A(\bx,t) = A^{(0)}(\bx,t)$, $n(\bk,\bx,t) = \eps^D n^{(0)}(\eps \bk, \eps \bx, t)$ described in the main text is the correct one to describe a hidden soliton stabilized by the IS.
%\begin{eqnarray}
%i\partial_t A  = - \frac{\alpha}{2} \partial_x^2 A + A V , 
%\label{eq:nls_c}\\
%\partial_t n({x},{k},t) +\alpha {k}  \partial_{{x}} n({x},{k},t) - \partial_{x} V({x},t)  \partial_{k} n({x},{k},t) = 0,
%\label{eq:vlasov}\\
%V({x},t) = - \gamma \int U({x}-{x}') \ \big( |A|^2({x}',t) + \rho({x}',t) \big) \ d{x}', \\
%\rho({x},t)=\frac{1}{2 \pi}  \int n({x},{k},t) d \bk.
%\end{eqnarray}
We look for solutions of the form
\begin{eqnarray}
\label{eq:sclaingappB}
n(\bk,\bx,t) = \eps^p n^{(0)}( \eps^q \bk,\eps^r \bx, \eps^s t), \quad \quad
A(\bx,t) = \eps^v A^{(0)}(\eps^w \bx,\eps^z t) ,
\end{eqnarray}
where $\eps\ll 1$ is a small dimensionless quantity that characterizes the scaling ratios
between the different characteristic length scales and amplitudes of the coherent and incoherent fields. 
With (\ref{eq:sclaingappB}) we also have $\rho_{IS}(\bx,t) = \eps^{p-D q} \rho_{IS}^{(0)}(\eps^r \bx , \eps^s t)$ with
$\rho_{IS}^{(0)}(\bX,T)= \int n^{(0)}(\bK,\bX,T) d\bK$.
We look for solutions with $r>w$, i.e., we look for a soliton whose radius is small compared to the typical radius of the IS. \\
%On souhaite aussi trouver une solution dans laquelle la structure coh\'erente est stabilis\'ee par la structure incoh\'erente.\\

I) We first look at the equations at the scale of the soliton. If $\bx = \eps^{-w} \bX$,
$t=\eps^{-z} T$, then
\begin{align*}
 (U_D* |A|^2)  A(\bx,t)&=  \eps^{3v-2w} (U_D*  |A^{(0)}|^2) A^{(0)}(\bX,T),\\
\Delta_\bx A (\bx,t) &= \eps^{v+2w} \Delta_\bX A^{(0)}(\bX,T) ,\\
\partial_t A (\bx,t) &= \eps^{v+z} \partial_T A^{(0)}(\bX,T) .
\end{align*}
We also have
\begin{align*}
 (U_D *  \rho_{IS}) (\bx,t)&=
%  \eps^{p-D q-2w} \int U_D( \bX-\bY) \rho_{IS}^{(0)}(\eps^{r-w} \bY,T) d\bY  = 
  \eps^{p-Dq-2r} \int U_D(\bY) \rho_{IS}^{(0)}(\bY+\eps^{r-w} \bX,T) d\bY .
\end{align*}
Since $r>w$ we can expand
%\begin{align*}
%\int U_D(\bY) \rho_{IS}^{(0)}(\bY+\eps^{r-w} \bX,T) d\bY =& 
% \int U_D(\bY) \rho_{IS}^{(0)}(\bY,\eps^{z-s} T)  d\bY \\
%&+ 
%\eps^{r-w} \bX \cdot \Big[ \int U_D(\bY)  \nabla_\bY \rho_{IS}^{(0)}(\bY,\eps^{z-s} T)  d\bY \Big]\\
%&+ 
%\frac{1}{2} \eps^{2r-2w} \bX \cdot \Big[ \int U_D(\bY) \nabla_\bY \otimes \nabla_\bY \rho_{IS}^{(0)}(\bY,\eps^{z-s} T)  d\bY \Big]\bX
%+ o( \eps^{2r-2w}).
%\end{align*}
\begin{align*}
\int U_D(\bY) \rho_{IS}^{(0)}(\bY+\eps^{r-w} \bX,T) d\bY =& 
 \int U_D(\bY) \rho_{IS}^{(0)}(\bY,\eps^{z-s} T)  d\bY 
+ 
\eps^{r-w} \bX \cdot \Big[ \int U_D(\bY)  \nabla_\bY \rho_{IS}^{(0)}(\bY,\eps^{z-s} T)  d\bY \Big]\\
&+ 
\frac{1}{2} \eps^{2r-2w} \bX \cdot \Big[ \int U_D(\bY) \nabla_\bY \otimes \nabla_\bY \rho_{IS}^{(0)}(\bY,\eps^{z-s} T)  d\bY \Big]\bX
+ o( \eps^{2r-2w}).
\end{align*}
The first term is a constant in $\bX$ that depends only on $T$. 
If $\rho_{IS}^{(0)}$ is spherically symmetric then the second term is zero and the third term takes the form
\begin{align*}
\int U_D(\bY) \rho_{IS}^{(0)}(\bY+\eps^{r-w} \bX,T) d\bY =& 
const_\eps
+\frac{1}{2 D} \eps^{2r-2w}  \Big[ \int U_D(\bY) \Delta_\bY \rho_{IS}^{(0)}(\bY,\eps^{z-s} T)  d\bY \Big] |\bX|^2
+ o( \eps^{2r-2w}).
\end{align*}
After integrating by parts and using $\Delta_\bY U_D(\bY)= -\eta_D \delta(\bY)$, 
we get
\begin{align}
\label{eq:intappB}
\int U_D(\bY) \rho_{IS}^{(0)}(\bY+\eps^{r-w} \bX,T) d\bY&= 
const_\eps
- q_D \eps^{2r-2w}  \rho_{IS}^{(0)}({\bf 0},\eps^{z-s} T)  |\bX|^2
+ o( \eps^{2r-2w}),
\end{align}
where $q_D = \eta_D /(2D)$,
and therefore
\begin{align*}
 (U_D *  \rho_{IS})A (\bx,t)&= 
const_\eps A^{(0)}(\bX,T) - q_D \eps^{v+p-Dq-2w} \rho_{IS}^{(0)}({\bf 0},\eps^{z-s} T) |\bX|^2  A^{(0)}(\bX,T) +o(\eps^{v+p-Dq-2w}).
\end{align*}
The term $ const_\eps A^{(0)}$ plays no role because it only gives a time-dependent phase term in (\ref{eq:nls_0}).
We require the $\eps$-terms to be balanced in (\ref{eq:nls_0}) because we look for a solution in the form of a soliton stabilized by the IS.
The terms are balanced 
if $ 3v-2w=v+p-Dq-2w=v+2w=v+z$, that is to say, if
\begin{align}
\label{eq:appBcond1}
(p-Dq)/2=2w=v=z .
\end{align}

II) We next look at the equations at the scale of the IS. If  $\bx = \eps^{-r} \bX$,
$\bk=\eps^{-q} \bK$, $t=\eps^{-s} T$, then (using $r>w$)
\begin{align*}
( U_D * |A|^2)(\bx,t)  &=  \eps^{2v -r(2-D)-wD} U_D(\bX) \int  |A^{(0)}(\bX',\eps^{s-z}T )|^2 d\bX' +o(\eps^{2v -r(2-D)-wD}) ,\\
(U_D * \rho_{IS})(\bx,t)&= \eps^{p-Dq-2r} U_D * \rho_{IS}^{(0)}(\bX,  T) .
\end{align*}
Thus
\begin{align*}
 \partial_{\bx} V  (\bx,t) \cdot \partial_{\bk} n (\bk,\bx,t)&=
\eps^{2p-(D-1)q-r} \partial_\bX \big(  U_D * \rho_{IS}^{(0)} \big) (\bX,T)\cdot \partial_\bK n^{(0)}(\bK,\bX,T)\\
&\quad +
\eps^{p+q+2v-r(1-D)-wD} \partial_\bX\Big( U_D(\bX) \int |A^{(0)}(\bX',\eps^{s-z}T )|^2 d\bX'\Big) \cdot \partial_\bK n^{(0)}(\bK,\bX,T) ,\\
{\bk} \cdot  \partial_{{\bx}} n (\bk,\bx,t)  &= \eps^{p-q+r} \bK \cdot \partial_\bX n^{(0)}  (\bK,\bX,t), \\
\partial_t n  (\bk,\bx,t) &= \eps^{p+s} \partial_T n^{(0)} (\bK,\bX,t) .
\end{align*}
We look for balanced terms in (\ref{eq:vlasov_0})  amongst the components coming the IS,
because we do not look for a solution in which the IS would be affected by the soliton.
The terms are balanced provided 
$2p-(D-1)q - r=p-q+r=p+s $, namely
%that is to say, if 
\begin{align}
\label{eq:appBcond2}
r=(p-(D-2)q)/2 \quad \mbox{ and } \quad s =(p-Dq)/2 .
\end{align}

III) Conclusion:
Without loss of generality, we can choose the reference length scale to be the radius of the soliton, that is to say, we can choose $w=0$.
Then we get balanced terms in equations in  (\ref{eq:nls_0})  and  (\ref{eq:vlasov_0})  provided conditions (\ref{eq:appBcond1}) and (\ref{eq:appBcond2})
are satisfied, which means
\begin{align}
\label{eq:appBcond}
w=v=s=z=0, \quad q=r >0 ,\quad p=Dq .
\end{align}
In this case we can check that $p+q+2v-r(1-D)-wD  =2Dq > (2D-1) q=  2p-r$,  which shows that the soliton has no influence on the IS:
%\begin{eqnarray}
\begin{align*}
\partial_{\bx} V  \cdot \partial_{\bk} n (\bk,\bx,t) &=
\eps^{Dq} \partial_\bX \big(  U_D * \rho_{IS}^{(0)} \big) (\bX,T)\cdot \partial_\bK n^{(0)}(\bK,\bX,T)
+o (\eps^{Dq}).
\end{align*}
%\end{eqnarray}
We can take without loss of generality $q=r=1$ and $p=D$, because the small dimensionless parameter $\eps$ is arbitrary.
We then obtain $A(\bx,t) = A^{(0)}(\bx,t)$, $n(\bk,\bx,t) = \eps^D n^{(0)}(\eps \bk, \eps \bx, t)$,
the SPE (\ref{eq:nls_0}) takes the form
\begin{eqnarray}
i\partial_t A^{(0)} (\bx,t)  = - \frac{\alpha}{2} \nabla^2 A^{(0)}(\bx,t) -\gamma (U_D*|A^{(0)}|^2)  A^{(0)}(\bx)+
 \gamma  q_D   \rho_{IS}^{(0)}({\bf 0},t) |\bx|^2 A^{(0)}(\bx),
\label{eq:nls_2}
\end{eqnarray}
that is to say the effective SPE Eq.(6) (main text), and the WT-VPE (\ref{eq:vlasov_0}) takes the form (with $\bX=\eps \bx$ and $\bK=\eps \bk$)
\begin{eqnarray}
\partial_t n^{(0)}(\bK,\bX,t) +\alpha \bK \cdot \partial_{\bX} n^{(0)}(\bK,\bX,t) +\gamma  \partial_{\bX} 
(U_D* \rho_{IS}^{(0)}  )(\bX,t) \cdot \partial_{\bK} n^{(0)}(\bK,\bX,t) = 0.
\end{eqnarray}
%that is to say, eqs.~(\ref{eq:vlasov_N0}-\ref{eq:V_0_inc}).
This rescaled form of the WT-VPE does not depend on the coherent component, i.e., on the soliton dynamics.

\bigskip
\noindent 
{\bf Justification of the scale separations of the new regime Eq.(3):} The scaling $A(\bx,t) = A^{(0)}(\bx,t)$, $n(\bk,\bx,t) = \eps^D n^{(0)}(\eps \bk, \eps \bx, t)$ also implies all of the separation of spatial scales involved in the new regime reported in our work, namely $\tilde{\xi}=\xi/\Lambda =\varepsilon$, $\lambda_c/\Lambda =O(\varepsilon)$, $\ell / \Lambda = O(\varepsilon^{-1})$, $R_S/\Lambda =O(1)$, $|A|^2 \sim O(1)$, $n \sim \varepsilon^D$ and $\rho_{IS}\sim 1$. This
gives Eq.(3):
$$\lambda_c \sim \xi  \ll R_S \sim \Lambda \ll \ell \quad \quad {\rm and} \quad \quad \rho_S \sim \bar{\rho}_{IS}.$$ 
%that is to say, Eq. (3).

%Remark: Here we have looked for a situation in  which the CS is stabilized by the IS.
%We have found that, necessarily, the radius of the CS is much smaller than the radius of the IS and much larger than the correlation radius of the IS, and the typical amplitudes of $\rho_{IS}$ and $|A|^2$ are of the same order.
%If we had looked for a situation in which the CS is stabilized by itself, without interaction with the IS, then the amplitude of $|A|^2$ would be larger than the one of $\rho_{IS}$.

\section{III. Binary soliton: 3D dynamics and derivation of Eqs.(7-10)}

We study the dynamics of the binary soliton system for $D=1$ and $D=3$.
The Lagrangian of the effective SPE (\ref{eq:nls_2}) is
\begin{eqnarray}
&& {\cal L} = \int  \frac{i}{2} \big( A \partial_t A^* -\partial_t A  A^*  \big)
+\frac{\alpha}{2}|\nabla A|^2 +\frac{1}{2} V_S(\bx) |A|^2  + q_D \gamma \rho_0 |\bx|^2 |A|^2  d\bx ,
\end{eqnarray}
where 
$V_S(\bx,t)= - \gamma \int U_D(\bx-\by) |A|^2(\by,t) d\by$, with $q_1=1$, $q_3=2\pi/3$, $U_1(x)=-|x|$, $U_3(\bx)=1/|\bx|$, and $\rho_0(t)=\rho_{IS}(\bx={\bf 0},t)$
is the average density of the  incoherent structure at the center. 
We consider the Gaussian ansatz for a two-component soliton:
\begin{eqnarray}
\nonumber
A(\bx,t) &=& \sum_{j=1}^2 a_j(t) \exp\Big( -\frac{|\bx-\bx_{o,j}(t)|^2}{2R_j^2(t)} +i b_j(t)|\bx-\bx_{o,j}(t)|^2 + i \bk_{o,j}(t)\cdot (\bx-\bx_{o,j}(t)) +i\nu_j(t)\Big) .
\label{eq:ansatz:2}
\end{eqnarray}
The effective Lagrangian depends on
$a_j(t)$, $R_j(t)$, $b_j(t)$, $\nu_j(t)$, $\bx_{o,j}(t)$ and $\bk_{o,j}(t)$ and  their  time  derivatives.
It can be split into three parts. The first two parts depend on each of the components of the coherent structure, the third part represents the interaction between the two components ${\cal L}  = {\cal L}_{1} + {\cal L}_{2} +{\cal L}_{12}$:
\begin{eqnarray*}
\nonumber
&&{\cal L}_{j} = M_{{S},j}\Big(\partial_t \nu_j+D\partial_t b_j \frac{R_j^2}{2} -\bk_{o,j} \cdot \partial_t \bx_{o,j}  \Big) +\frac{\alpha M_{{S},j}}{4}
\big(\frac{D}{R_j^2}+ 4D b_j^2 R_j^2 + 2|\bk_{o,j}|^2 \big) \nonumber \\
&& \quad \quad - \frac{\gamma}{\sqrt{2\pi}}M_{S,j}^2R_j^{2-D} + \gamma \rho_0 q_D  M_{{S},j}\big(\frac{D}{2} R_j^2+ |\bx_{o,j}|^2\big)  ,\\
&&{\cal L}_{12} = - \gamma M_{{S},1}M_{{S},2}
|\bx_{o,1}-\bx_{o,2}|^{2-D},
\end{eqnarray*}
where $M_{S,j} =a_j^2 \pi^{D/2} R_j^D$ is the mass of the $j$-th component of the coherent structure
and we have assumed that $|\bx_{o,1}-\bx_{o,2}|\gg R_1,R_2$ in order to simplify ${\cal L}_{12}$.
The  evolution  equations  for  the
parameters of the ansatz are then derived from the effective
Lagrangian   by   using   the   corresponding   Euler-Lagrange
equations $\delta \int {\cal L} dt =0$ \cite{malomed02}. 
We obtain the closed-form ordinary differential equation for the width $R_j$ of the $j$-th component of the coherent structure:
\begin{equation}
\label{eq:rayon_1d}
\partial_t^2 R_j = \frac{\alpha^2}{R_j^3} - \sqrt{\frac{2}{\pi}} \frac{\alpha \gamma M_{S,j}}{D R_j^{D-1}} - 2 q_D \alpha  \gamma \rho_0 R_j.
\end{equation}
We also obtain  the closed-form and coupled system of ordinary differential equations for the centers $\bx_{o,j}$ and central wavenumber $\bk_{o,j}$ of the coherent structure:
\begin{eqnarray}
\label{eq:center_1d}
&&\partial_t \bx_{o,j} =  \alpha \bk_{o,j} , \quad j=1,2, \\
&&\partial_t \bk_{o,j} =  -2q_D  \gamma \rho_0 \bx_{o,j} - \gamma M_{S,3-j}  \frac{\bx_{o,j}-\bx_{o,3-j}}{|\bx_{o,j}-\bx_{o,3-j}|^D} , \quad j=1,2.
\label{eq:kappa_1d:2}
\end{eqnarray}

\subsection{Mass-radius relation for the hidden soliton}

Let us first examine Eq.(\ref{eq:rayon_1d}) for the radius $R_j(t)$ of the $j$-th soliton component.
The energy is of the form
$ 
\frac{1}{2} (\partial_t R_j)^2 + W_j(R_j) ,
$
with the effective potential
$$
W_j(R_j) = \frac{\alpha^2}{2 R_j^2} + \sqrt{\frac{2}{\pi}} \frac{\alpha \gamma M_{S,j}}{D(D-2) R_j^{D-2}} + q_D \alpha\gamma \rho_0 R_j^2 .
$$
Equation (\ref{eq:rayon_1d}) has a stable equilibrium $R_j=R_{S,j}$ provided $\partial_{R_j} W_j(R_{S,j})=0$ and $\partial_{R_j}^2 W_j(R_{S,j})>0$.
For any positive mass $M_{S,j}$, there is a unique stable solution with radius $R_{S,j}(M_{S,j})$ that is the unique solution to the quartic equation
\begin{equation}
M_{S,j} = \sqrt{2\pi} D R_{S,j}^{D} {\bar \rho} \big( (\Lambda/R_{S,j})^{4}-q_D \rho_0/{\bar \rho} \big).
\label{eq:masse_1d}
\end{equation}

\subsection{Motion of the center of mass of the binary soliton}

\subsubsection{Orbital revolution period $\tau_{\rm cm}$}

The soliton barycenter 
$
\bX_{\rm cm} = (M_{S,1} \bx_{o,1}+M_{S,2} \bx_{o,2})/(
M_{S,1}+M_{S,2}) $,
%\quad \quad 
$\bK_{\rm cm} = (M_{S,1} \bk_{o,1}+M_{S,2} \bk_{o,2})/(M_{S,1}+M_{S,2} )$
satisfies
\begin{eqnarray}
&&\partial_t \bX_{\rm cm} = \partial_{\bK_o} H_{\rm cm} =  \alpha \bK_{\rm cm} , \\
&&\partial_t \bK_{\rm cm} = -\partial_{\bX_o} H_{\rm cm} = - 2 q_D \gamma \rho_0 \bX_{\rm cm},
\end{eqnarray}
with the conserved Hamiltonian $H_{\rm cm}=q_D \gamma \rho_0 |\bX_{\rm cm}|^2 +\frac{\alpha}{2} |\bK_{\rm cm}|^2$.
The coherent structure barycenter then exhibits a periodic ellipsoidal motion in phase-space.
The revolution period is
$\tau_{\rm cm}= \sqrt{2} \pi/\sqrt{q_D \alpha \gamma \rho_0}$.
%If the initial state is such that $X_{o}(0)=0$ and $K(0)=0$, then the CS is such that $X_{o}(t)=0$ and $K(t)=0$ for all $t$.

\subsubsection{The case $D=3$: Ellipsoidal motion}

For $D=3$, the motion of $\bX_{\rm cm}$ lies in a plane (spanned by the initial conditions $\bX_{\rm cm}(0)$ and $\partial_t \bX_{\rm cm}(0)$) and follows an ellipse.
In the plane of the trajectory, the motion of $\bX_{\rm cm}$ has the form
$
\bX_{\rm cm} = ({\cal R}(\theta) \cos \theta, {\cal R}(\theta) \sin \theta, 0) ,
$
where 
$$
{\cal R}(\theta) = \frac{c_1^{-1/4}}{\sqrt{w_- \cos^2\theta+w_+\sin^2 \theta}} ,\quad \quad w_\pm=C\pm\sqrt{C^2-1}, \quad \quad \theta(t) = {\rm arctan}\Big( w_- \tan( c_o \sqrt{c_1} t)\Big),
$$
$c_o = \partial_t  \bX_{\rm cm} \cdot \bX_{\rm cm}^\perp$ is a constant of motion, $ \bX_{\rm cm}^\perp=   (- {\cal R}(\theta) \sin \theta, {\cal R}(\theta) \cos \theta, 0)$,
$$
C= \frac{1}{2 \sqrt{c_1} |\bX_{\rm cm}|^2}+\frac{\sqrt{c_1} |\bX_{\rm cm}|^2}{2}+\frac{1}{2\sqrt{c_1} c_o^2 } \frac{(\bX_{\rm cm} \cdot \partial_t \bX_{\rm cm})^2}{|\bX_{\rm cm}|^2}
$$ 
is a constant of motion, and
$
c_1= \frac{4\pi \alpha\gamma \rho_0}{3c_o^2}.
$
The motion follows an ellipse:
$$
w_- X_{\rm cm,1}^2
+
w_+ X_{\rm cm,2}^2
=c_1^{-1/2},
$$
where we have assumed that the radius is maximal at time $0$ and $\theta(t=0)=0$.

\subsection{Relative motion of the binary soliton}

The relative motion of the binary soliton is defined from $\bX_o = \bx_{o,1}-\bx_{o,2}$, $\bK_o = \bk_{o,1}-\bk_{o,2}$, which satisfies
\begin{eqnarray*}
&&\partial_t \bX_o = \partial_{\bK_o} H_o = \alpha \bK_o , \\
&&\partial_t \bK_o = -\partial_{\bX_o} H_o = - 2 q_D \gamma \rho_0 \bX_o -\gamma (M_{S,1} +M_{S,2} ) 
\bX_o / |\bX_o|^{D},
\end{eqnarray*}
with the conserved Hamiltonian 
%\begin{equation*}
$H_{o}= q_D\gamma \rho_0 |\bX_o|^2 + \gamma (M_{S,1} +M_{S,2} ) 
|\bX_o|^{2-D} + \alpha |\bK_o|^2/2$. 
%\end{equation*}

\begin{figure}
\includegraphics[width=.23\columnwidth]{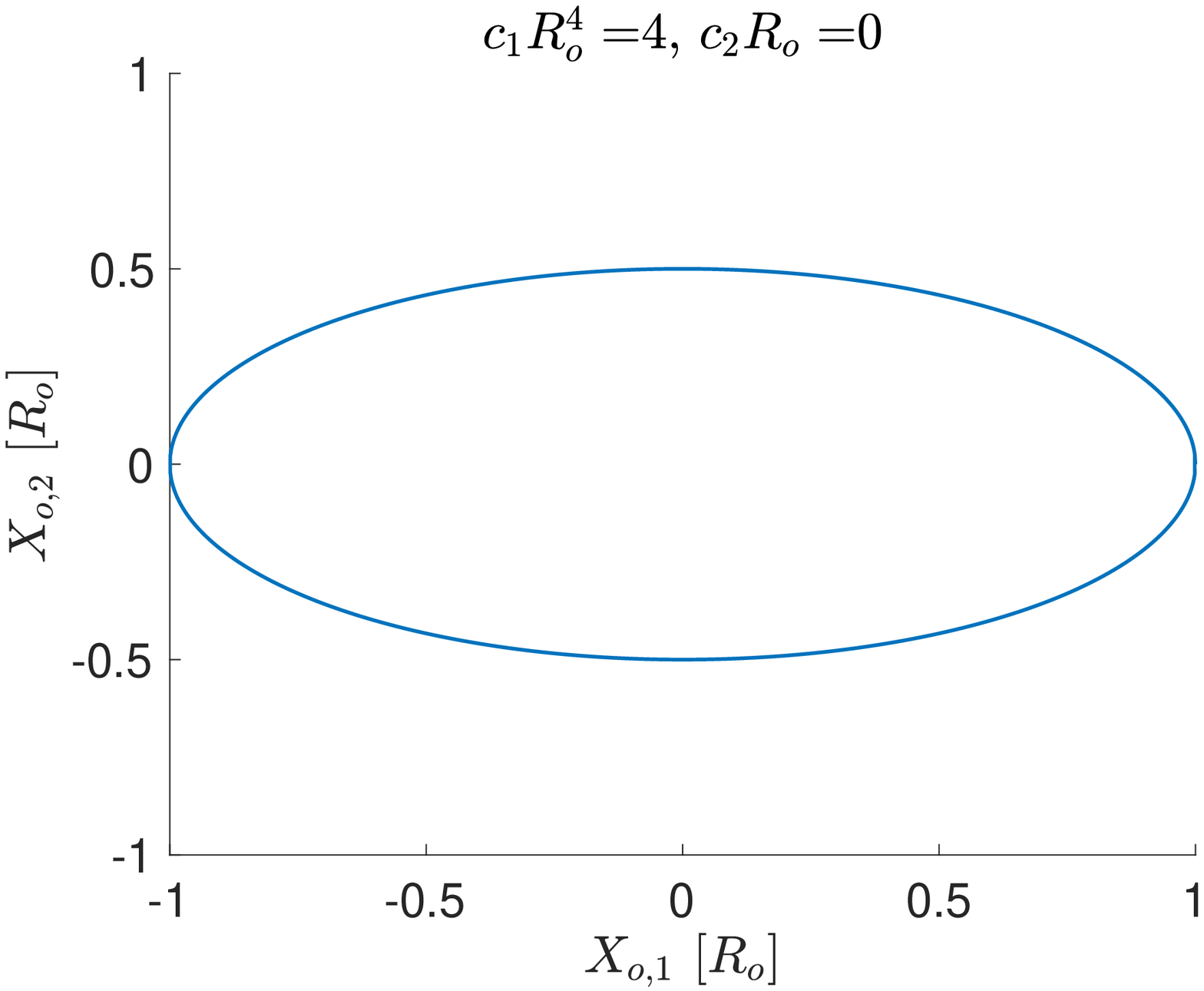}
\includegraphics[width=.23\columnwidth]{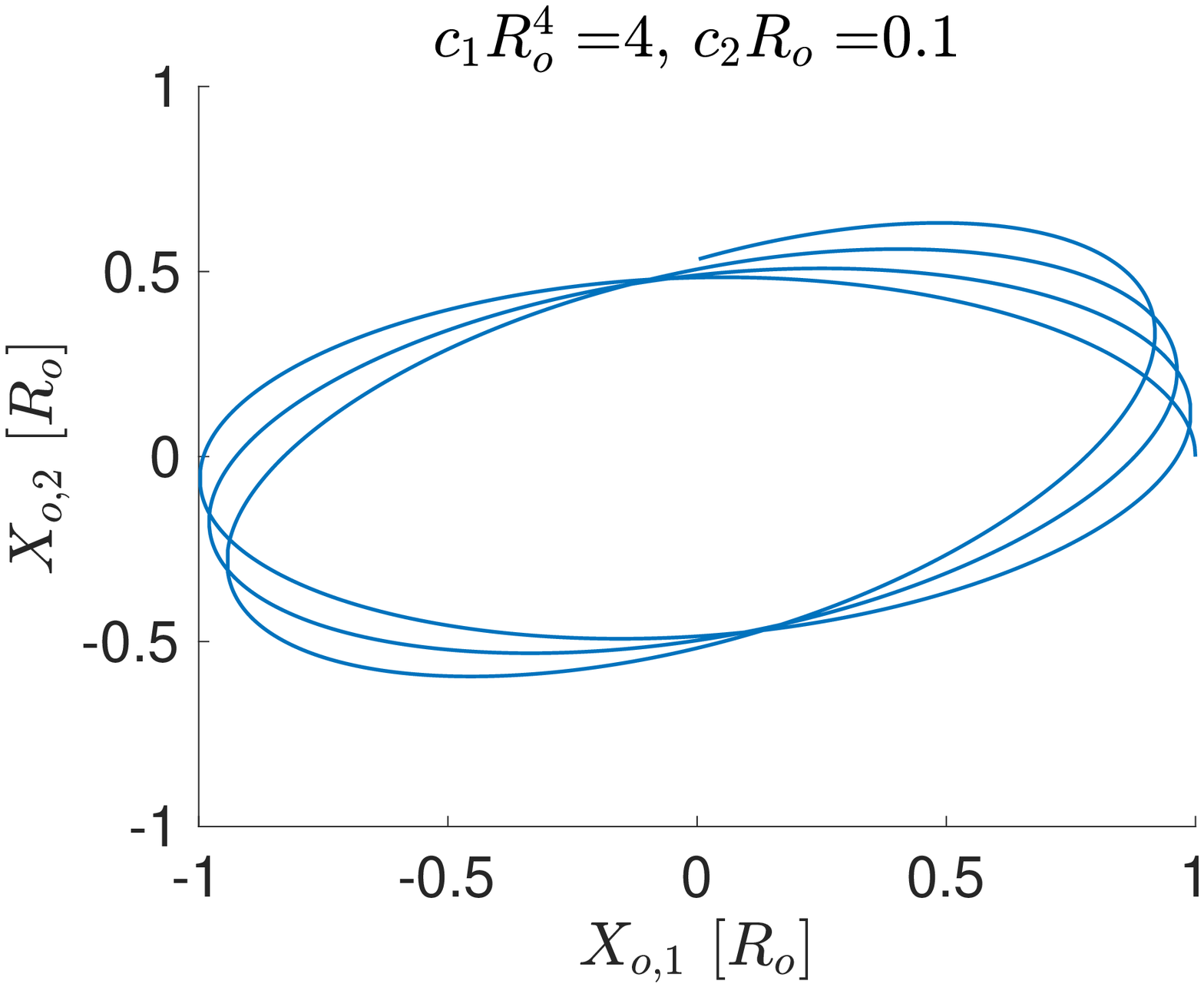}
\includegraphics[width=.23\columnwidth]{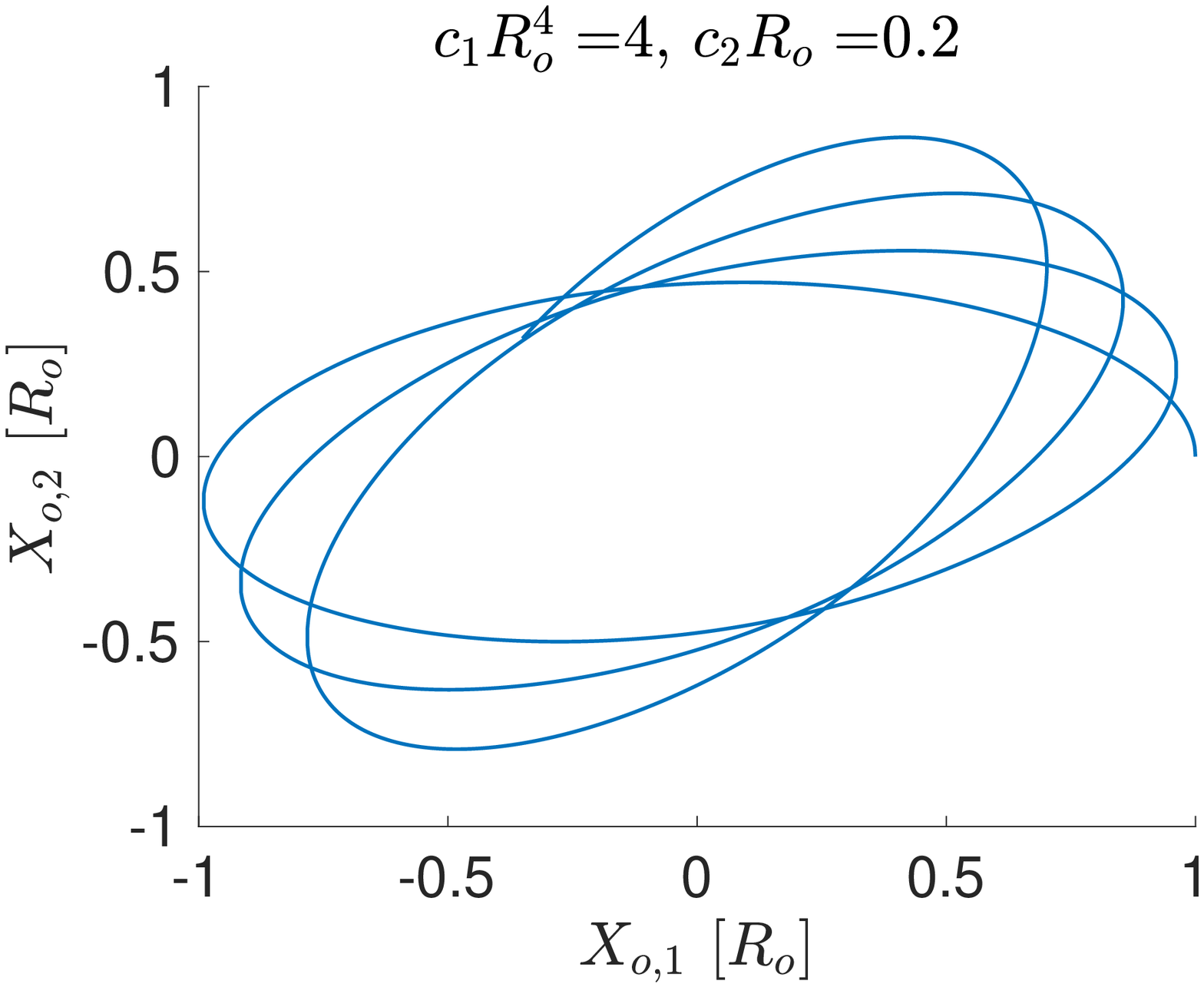}
\includegraphics[width=.23\columnwidth]{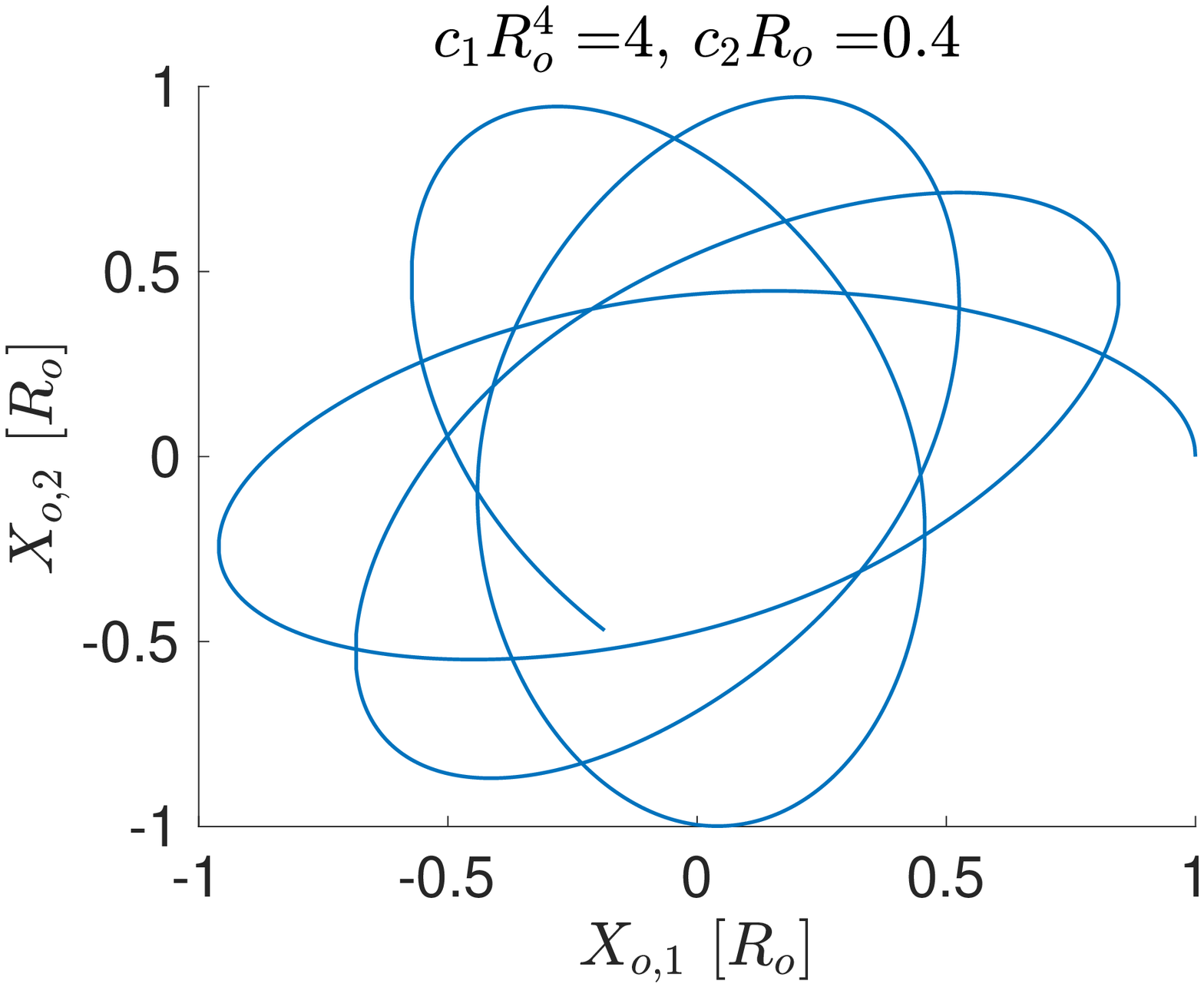}
\caption{
\baselineskip 8pt
{\bf Perihelion precession of the 3D binary soliton:} Planar trajectory of the relative motion of the bisoliton $\bX_o(t)$ for $t \in [0,10 \tau_o]$, $\bX_o(0)=(R_o,0,0)$, $\partial_t \bX_o(0) \cdot \bX_o(0)=0$,
$c_o = \partial_t \bX_o(0) \cdot \bX_o(0)^\perp =R_o^2 / \tau_o$,
and for different values of $(c_1,c_2)$.
}
\label{fig:precession} 
\end{figure}

\subsubsection{The case $D=1$: Revolution period $\tau_{\rm bin}$ of the binary soliton}

The trajectory has the form of two half ellipses connected through two points that present a small cusp.
The relative motion of the binary soliton is then periodic in phase-space with a revolution period 
$$
\tau_{\rm bin}=\frac{2 \sqrt{2}}{\sqrt{\alpha}} \int_{0}^{\xi(E)} \frac{dx}{\sqrt{E-U(x)}} ,
$$
with $U(x) = \gamma \rho_0 x^2+\gamma (M_{S,1} +M_{S,2} ) x$ and 
$\xi (E) = \big(-\gamma(M_{S,1} +M_{S,2} )  +\sqrt{\gamma^2 (M_{S,1} +M_{S,2} ) ^2 +4 \gamma \rho_0 E}\big)/(2 \gamma \rho_0)$.
The integration gives Eq.(10) (main text).
%, namely
%$$
%{\tau}_{\rm bin} =
%\frac{4\sqrt{2}}{\sqrt{\alpha \gamma \rho_0}} {\rm arcsin}
%\Big( \sqrt{\frac{1}{2} - \frac{\gamma (M_{S,1} +M_{S,2} ) }{2\sqrt{\gamma^2 (M_{S,1} +M_{S,2} ) ^2 +4\gamma \rho_0 H_o}}}\Big).
%$$
%%Note that we always have ${\tau}_{\rm bin} \leq \tau_{\rm cm}$.
%For instance, a two-component coherent structure with $M_{S,1}=M_{S,2}$ and $X_o=0$, $K=0$, satisfies $x_{o,1}(t)=-x_{o,2}(t)$ and $\kappa_1(t)=-\kappa_2(t)$ and the motion is faster than for a single-component structure.
%However, we typically have $\rho_0 \sim M_S /R_s$ and $R_s \ll |\Delta X_o|$, so $\rho_0 |\Delta X_o| \gg M_S$ and the correction term due to the interaction is small.
%For instance the period can be expanded as 
%$$
%{\tau}_{per,2} 
%\simeq
% \tau_{per}  \Big[1 - 
% \frac{\sqrt{\gamma} (M_{S,1} +M_{S,2} ) }{\pi \sqrt{\rho_0 E}
% }
% \Big].
%$$

\subsubsection{The case $D=3$: Motion of the binary soliton}

The motion of $\bX_o(t)$ lies in a plane. In the plane of the trajectory, 
the motion of $\bX_o$ has the form
$
\bX_o = ({\cal R}(\theta) \cos \theta, {\cal R}(\theta) \sin \theta, 0)  ,
$
where $u(\theta) = 1/{\cal R}(\theta)$ is solution of
$
\partial_\theta^2 u +u = \frac{c_1}{u^3}+c_2 ,
$
and $\theta(t)$ is solution of 
$
\partial_t \theta = \frac{c_o}{{\cal R}(\theta)^2}  ,
$
where $c_o = \partial_t \bX_o \cdot \bX_o^\perp$ is a constant of motion, $\bX_o^\perp=   (- {\cal R}(\theta) \sin \theta, {\cal R}(\theta) \cos \theta, 0)$,
$$
C= \frac{1}{2 \sqrt{c_1} |\bX_o|^2}+\frac{\sqrt{c_1} |\bX_o|^2}{2}+\frac{1}{2\sqrt{c_1} c_o^2 } \frac{(\bX_o \cdot \partial_t \bX_o)^2}{|\bX_o|^2} - \frac{c_2}{\sqrt{c_1} |\bX_o|}
$$ 
is a constant of motion, and
$c_1= 2 q_D  \alpha\gamma \rho_0/c_o^2$, $c_2 = \alpha \gamma (M_{{\rm S},1} +M_{{\rm S},2} ) /c_o^2$.
Note that the function $u$ or ${\cal R}$ is periodic in $\theta$, but its period is not $2\pi$ (it is $2\pi$ if $c_1=0$ or $c_2 = 0$). 
%That is why the orbit of $\bX_o$  is not closed, see Fig.~\ref{fig:precession}.
Then the relative motion of the binary soliton is not a closed orbit and it is not periodic, except for some exceptional values of the parameters, see Fig.~\ref{fig:precession}.

\subsubsection{Computation of the revolution periods of the single soliton and binary soliton in SPE simulations}

The evaluation in the simulations of the revolution period of the single soliton in Fig.~3 and of the center of mass of the binary soliton in Fig.~2 requires the computation of the density $\rho_0$ of the incoherent structure, as well as the mass(es) of the soliton(s) $M_{S,(j)}$.
The soliton mass is computed in phase-space from the Husimi function $W(k,x,t)$.
In the case of Fig.~3, we obtain $M_S/{M} = 0.039 \pm 0.002$; in the case of Fig.~2, $M_{S,1}/{M} =0.037\pm 0.002$, $M_{S,2}/{M} =0.025 \pm 0.002$.
The uncertainties on the masses are determined from the variance of the  fluctuations over the relevant time interval.
$\rho_0$ is retrieved by computing the spatial and temporal averages of $\left< |\psi(x,t)|^2 \right>$ over a small spatial window $\Delta x_w \simeq 7.5 \Lambda$ and the relevant time interval.
In order to account solely for the contribution of the IS component, we have removed the contribution of the soliton mass(es) $M_{S,(j)}$ to the computation of $\rho_0$.
In the case of Fig.~3, we obtain $\rho_0/{\bar \rho}= 4.2 \pm 0.2$; in Fig.~2 $\rho_0/{\bar \rho}= 3.9 \pm 0.2$ (the uncertainties being determined from the variance of the  fluctuations).
This gives for the single soliton of Fig.~3 $\tau_{\rm cm}/\tau = 1.52 \pm 0.04$; and for the  binary soliton of Fig.~2 $\tau_{\rm cm}/\tau = 1.56 \pm 0.04$, and $\tau_{\rm bin}/\tau=1.43 \pm 0.04$.
These values are in agreement with those observed in the SPE simulations (see the main text: $\tau_{\rm cm}^{\rm num}/\tau = 1.52$ for Fig.~3; $\tau_{\rm cm}^{\rm num}/\tau = 1.56$, and $\tau_{\rm bin}^{\rm num}/\tau=1.43$ for Fig.~2).
%Note that the particular choice of the size of the spatial window $\Delta x_w$ does not affect the validity such a good agreement.

\end{widetext}

%\newpage


\end{document}